\documentclass[12pt,preprint]{aastex}

\begin{document}

\title{A Puzzle Involving Galactic Bulge Microlensing Events 
\altaffilmark{1}}

\author{Judith G. Cohen\altaffilmark{2},  Andrew Gould\altaffilmark{3},
Ian B. Thompson\altaffilmark{4}, Sofia Feltzing\altaffilmark{5}, 
Thomas Bensby\altaffilmark{6},
Jennifer A. Johnson\altaffilmark{3}, Wenjin Huang\altaffilmark{7},
Jorge Mel\'endez\altaffilmark{8}, Sara Lucatello\altaffilmark{9}   \& 
Martin Asplund\altaffilmark{10} }

\altaffiltext{1}{Based in part on observations obtained at the
W.M. Keck Observatory, which is operated jointly by the California 
Institute of Technology, the University of California, and the
National Aeronautics and Space Administration.}

\altaffiltext{2}{Palomar Observatory, Mail Stop 249-17,
California Institute of Technology, Pasadena, Ca., 91125, 
jlc@astro.caltech.edu}

\altaffiltext{3}{Department of Astronomy, Ohio State
University, 140 W. 18th Ave., Columbus, OH 43210; 
gould,jaj@astronomy.ohio-state.edu}

\altaffiltext{4}{Carnegie Observatories,
813 Santa Barbara Street, Pasadena, Ca. 91101,
ian@obs.carnegiescience.edu}

\altaffiltext{4}{Lund Observatory, Box 43, SE-22100 Lund, Sweden,
sofia@astro.lu.se}

\altaffiltext{6}{European Southern Observatory, Alonso de Cordova 3107,
Vitacura, Casilla 19001, Santiago 19, Chile, tbensby@eso.org}

\altaffiltext{7}{Palomar Observatory, Mail Stop 105-24,
California Institute of Technology, Pasadena, Ca., 91125,
current address: University of
Washington, Department of Astronomy, Box 351580, 
Seattle, Washington, 98195-1580, hwenjin@astro.washington.edu}

\altaffiltext{8}
{Centro de Astrof\'{\i}sica da Universidade do Porto, Rua das Estrelas, 4150-762 
Porto, Portugal (jorge@astro.up.pt) }

\altaffiltext{8}{INAF-Astronomical Observatory of Padova,
Vicolo dell'Osservatorio 5,
35122 Padova, Italy,
and
Excellence Cluster Universe, Technische Universität München, Boltzmannstr. 2
D-85748 Garching, Germany, sara.lucatello@oapd.inaf.it}

\altaffiltext{10}{Max Planck Institute for Astrophysics,
Postfach 1317, 85741 Garching, Germany, asplund@MPA-Garching.MPG.DE}

\begin{abstract}
We study a sample of 16 microlensed Galactic bulge main
sequence turnoff region stars for which high dispersion
spectra have been obtained with detailed abundance analyses.
We demonstrate that there is a very
strong and highly statistically significant correlation 
between the maximum magnification
of the microlensed bulge star and the value of 
the [Fe/H] deduced from the high resolution
spectrum of each object. 
Physics demands that this correlation, assuming it to be real, be the result
of some sample bias.  We suggest several possible
explanations, but are forced to reject them all,
and are left puzzled.  
To obtain a reliable metallicity distribution
in the Galactic bulge 
based on microlensed dwarf stars it will
be necessary to resolve this issue through the course of
additional observations.

\end{abstract}


\section{Introduction \label{section_intro} }

Microlensing occurs when a ``lens'' (star, planet, black hole, etc)
becomes closely aligned with a more distant ``source'' star, whose image
it both magnifies and distorts.  Microlensing of stars in the
Galactic bulge offers the possibility of studying in detail
stars that are too faint for such even with the
largest existing telescopes without a microlensing boost.  
The lens
is usually  a star  the Galactic bulge, but sometimes in the disk of the 
Milky Way \citep{dominik06}.

Bulge giants are bright enough that 
high-dispersion spectra can routinely be obtained
 with 8~m class telescopes.
Extensive surveys of such giants 
at optical wavelengths \citep[see, e.g.][]{fulbright06} have
been carried out to construct the 
metallicity distribution function (MDF) for the bulge,
many of them in Baade's Window
($b \sim -4^{\circ}$),  the innermost field of relatively
low reddening.
\cite{zoccali08} have presented  results
of a survey of [Fe/H] in the Galactic bulge
from spectra  of about  800 stars with $\lambda/\Delta\lambda = 20,000$.
They find a radial gradient in [Fe/H] within the bulge
with the mean value going from +0.03~dex at $b = -4^{\circ}$ to
$-0.12$~dex at $b = -6^{\circ}$, and a sharp cutoff toward higher
metallicities with less than 5\% of the sample in Baade's Window having
[Fe/H] $> 0.4$~dex.  
\cite{rich07}, who reach into
($l,b) = (0^{\circ},-1^{\circ}$) with high dispersion in the near-IR, 
still find a sub-solar mean metallicity
of $-0.22\pm0.01$~dex.

The ability to obtain high resolution, high quality spectra
of microlensed Galactic bulge dwarfs and to carry out a detailed abundance
analysis offers an independent
apparently unbiased way to determine the
MDF of Galactic bulge stars, as well as their
detailed chemical inventory.
In principle the abundance analysis of
a upper main sequence dwarf is much easier and less prone to error
for spectra of a fixed signal-to-noise ratio
than that of a much cooler but much brighter bulge giant with a very
complex spectrum full of blends and strong molecular bands.

\section{The Sample of Microlensed Galactic Bulge Stars \label{section_sample} } 

We consider here the sample of subgiants near the main sequence
turnoff (MSTO), stars at the MSTO, or dwarfs of slightly
lower luminosity than the MSTO which have 
had high-magnification microlensing events during which
high-dispersion spectra have been obtained with large optical telescopes.  
We denote this
group as the microlensed MSTO dwarfs in the Galactic bulge.
Our sample includes all such stars with published abundance
analyses.  

The pioneering work of \cite{cavallo} includes
three such stars,
MACHO-98-BLG-6, MACHO-99-BLG-1\footnote{We omit this star as it's
spectrum shows double lines according to \cite{bensby10b}.}, and MACHO-99-BLG22;
see, e.g. \cite{stubbs93} and \cite{alcock99} for descriptions
of the MACHO survey.
\cite{johnson07} (MOA-2006-BLG265)
were the first to piggy-back on the microlensing planet hunters,
who prize high-magnification events because of their
extreme sensitivity to planets \citep{ob05071,ob05169,ob06109}.
This star
proved to be extremely metal-rich, 
much more so than the bulk of the much larger samples
of bulge giants, arousing considerable interest.
Improvements in the current
generation of microlensing surveys of the bulge,
the OGLE 
collaboration\footnote{http://www.astrouw.edu.pl/$\sim$ogle/ogle3/ews/ews.html} 
\citep{ews} and the MOA 
collaboration\footnote{http://www.phys.canterbury.ac.nz/moa} 
\citep{bond_moa}, led to increasing
numbers of alerts; they together find a total of about 800 microlensing
events per year, of which the Microlensing Follow Up
Network ($\mu$FUN)\footnote{http://www.astronomy.ohio-state.edu/$\sim$microfun/} 
is able to identify about 10 as
high-magnification events.

Recognition of the importance of observations of transient sources
has led to modifications in telescope  operations to enhance
our ability to take advantage of such brief opportunities.
Thus \cite{johnson08}
(MOA-2006-BLG99) and \cite{cohen08} (OGLE-2007-BLG-349S), the
latter spectrum with 
signal-to-noise ratio/spectral
resolution element (SNR) $>~$90
for $\lambda > 5500$~\AA,
rapidly followed, as did \cite{cohen09} 
(MOA-2008-BLG-310S and 311S), as well as the very recently
completed analysis of  OGLE-2007-BLG514S \citep{epstein10}.  Each of these 
microlensed MSTO bulge stars turned out to have
very high [Fe/H].
The first three of these led to 
the suggestion by \cite{cohen08} that the true MDF 
in the Galactic bulge was that of the dwarfs,
characterized by significantly higher mean [Fe/H] than
that of the giant bulge samples.

All of these stars were observed with either HIRES \citep{vogt94}
at the Keck I telescope or the MIKE spectrograph  \citep{bernstein03} 
at the 6.5~m Magellan Clay
Telescope at the Las Campanas Observatory.  
Last year, a group led by
S.~Feltzing, T.~Bensby and J.~Johnson began
observing MSTO microlensed bulge stars with UVES \citep{uves} at the VLT.
Results from two stars, OGLE-2008-BLG-209S \citep[][a MIKE spectrum]{bensby09a}
and OGLE-2009-BLG-076S \citep{bensby09b} followed,
both of which were metal-poor, with [Fe/H] $-0.33$
and $-0.76$~dex respectively.  \cite{bensby10b} suggested
that  the previous high metallicities for
microlensed bulge dwarfs might be just a matter
of chance; they find that they cannot reject
the possibility that the MDFs for the bulge giants and microlensed
dwarfs are identical.

Here we adopt the [Fe/H] and their associated
uncertainties given in the published papers
referenced above.  
To this sample we add MOA-2009-BLG259S, observed
with HIRES/Keck in July 2009 for which  Cohen et al.\ (in preparation) find
[Fe/H] = +0.55~$\pm0.10$~dex.  We also add five microlensed bulge dwarfs
observed with the VLT from \cite{bensby10b} for a total sample of 16 microlensed 
MSTO Galactic bulge stars.

\section{The Maximum Magnification of the Microlensing Events \label{section_amplitude} }

The maximum magnification, $A(max)$, of each of of these microlensing events
is based for all recent events on fits to high cadence photometry obtained by
$\mu$FUN\footnote{The $A(max)$ from the
MOA web site are preliminary values only.}.
\cite{macho_maxamp} give $A(max)$ for the MACHO events;
we use \cite{moa310_phot} for
MOA-2008-BLG-310S.
$A(max)$, the ratio of the apparent  brightness at the peak of 
the microlensing
event to that before or after, is independent of reddening.
It ranges from 5 to $\sim$1000 for our sample stars 
(see Table~\ref{table_data}).
 Figure~\ref{figure_max_mag}
shows the relationship between $A(max)$ for the event and the [Fe/H] 
of the source derived
from the high-dispersion spectra for each of the 16 microlensed Galactic
bulge stars in our sample.

It is apparent from Figure~\ref{figure_max_mag}
that there is a very strong correlation between 
$A(max)$ and  [Fe/H] for the MSTO sample of microlensed 
bulge stars. A similarly strong correlation is shown
when $A$ at the time the spectra were acquired is used instead.
A Spearman rank test indicates that the two-sided probability
that $A(max)$ is not correlated with [Fe/H] is
$5 \times 10^{-3}$. If MACHO-1998-BLG6, which \cite{bensby10b}
consider a low luminosity giant, not a subgiant, is omitted,
the two-sided probability for the remaining 15 stars becomes
$6 \times 10^{-3}$, not significantly larger.
While in principle this could be a statistical fluke, the
formal probability of this is sufficiently low to investigate
the implications of it being a real effect.

Figure~\ref{figure_on_sky} shows the locations on the
sky of the sample of microlensed bulge stars.  The region of
positive
Galactic latitude in general has significantly larger extinction than
that of negative $b$ in the region of the Galactic center, and much
of that area is not covered by any microlensing survey.
There is no obvious difference in the
projected spatial distribution on the sky of the very high vs 
the lower $A(max)$
bulge microlensed MSTO stars.
The maximum
projected distance from the Galactic center for a star in our sample is
6.6$^{\circ}$ (0.9~kpc), the minimum 2.2$^{\circ}$);  the median
is 4.0$^{\circ}$, the same as that
of Baade's Window.

Thus in addition to the discrepancy between the MDF
of the bulge giants vs that of the microlensed dwarfs
reviewed in \S\ref{section_intro}, 
an even more puzzling and potentially more serious
problem is introduced by Fig.~\ref{figure_max_mag}.
At first glance this figure suggests the presence
of two populations, one at [Fe/H] about +0.35~dex with 
$\sigma$([Fe/H]) small and with $A(max) > 200$, and one with
mean [Fe/H] considerably lower at about $-0.4$~dex, a larger
dispersion in metallicity, and with $A(max) < 200$. 
Since we have already shown that the
projected distributions of the low and high magnification
events are similar, we next consider
systematically different positions along the line of sight.
One might imagine that spatially
separating the high and low metallicity microlensing events,
having them arise in regions of different stellar density
(presumably closer or further from the Galactic center) 
in the presence of radial gradients in [Fe/H] 
within the bulge \citep[already established as present further out
in the bulge by][]{zoccali08} could
lead to the very strong correlation seen in Figure~\ref{figure_max_mag}.

Microlensing with stars as both sources and as lenses is a 
phenomenon depending only on geometry (i.e. distances and impact parameter) and
the mass of the lens.  An ensemble of microlensing events
also has properties that depend on the spatial density
of the sources and lenses. 
The distribution in magnification for a particular
source and a particular lens
always has a much higher probability for events with large 
impact parameter 
(low $A(max)$)
than it does for events with small impact parameter 
(high-magnification events).
The absence of any event 
with high [Fe/H] and $A(max) < 200$  given the many very
high $A(max)$, high [Fe/H] events we see
poses an insurmountable problem to any hypothesis
that seeks to explain the trend seen in Fig.~\ref{figure_max_mag}
through a mechanism of spatially separate populations within the bulge with
differing mean metallicities, densities, etc.

If the basic physical laws governing microlensing are not to
be violated, there must be a previously unrecognized sample bias
that produces the strong correlation 
between $A(max)$ and [Fe/H] seen in Fig.~\ref{figure_max_mag}.
We consider several possibilities below.

(1)  When the angular size of the
impact parameter for the microlensing event is so
small that it becomes comparable to the angular size of the source
star's radius, 
finite source effects, i.e. differential 
magnification of the limb relative to that of the center of the disk of
the source star, become important.  This occurs only in 
very high $A(max)$ events, and affects
 the strength of
spectral lines, which could potentially produce a spurious [Fe/H]
in the highest $A(max)$ events, leading to a systematic sample bias.
\cite{johnson10} 
carried out detailed abundance studies
of synthetic spectra of highly magnified dwarf stars.  They concluded
that the effects are always less than 0.05~dex, which is 
much too small to account
for the effect seen in Fig.~\ref{figure_max_mag}.

(2) Perhaps in the very crowded fields of the Galactic bulge,
the spectra in  low magnification events include
a substantial contribution from close neighbors of the microlensed
dwarf, while when $A(max)$ is large, spectroscopic observations
only detect the source star.  We have evaluated the contribution of
blending stars by comparing the unlensed brightness inferred from
the microlensing light curve with that from OGLE or MOA
images long before the event{\footnote{The required data are not
easily available for the MACHO events.}.  In the worst case (MOA-2009-BLG493), 
there is a 10\% contribution at the time of observation, the second
worst case has a 5\% contribution by blending stars; for all other
events this is not an issue.  This demonstrates
that blending is not the answer. Furthermore
double-lined spectra are not seen among any of these microlensed dwarfs.

(3) Perhaps for some
of the events the source is a foreground
disk star.   If disk stars  
have a  lower mean [Fe/H]  than that of the bulge
and if such events preferentially
include
those with low $A(max)$, this could reproduce the observed
relationship shown in Fig.~\ref{figure_max_mag}.
However, extrapolating the linear fit to the metallicity gradient
determined by \cite{luck06} 
outside 4~kpc from the center further inward, the disk
would reach [Fe/H] +0.5~dex at $R_{GC} = 1$~kpc.
In addition,
calculations of the probability of microlensing for
a source in the Galactic disk \citep[see e.g.][]{kane06} 
demonstrate that the source of a microlensing event
towards the bulge actually be a foreground disk star is  
unlikely.  Furthermore the measured
radial velocities show that the sources
of these microlensing events are a kinematically hot population
with ${\sigma}(v_r)$ much too large for the Galactic disk;
see Fig.~\ref{figure_on_sky}.  Additional arguments supporting
a bulge origin for the MSTO microlensing dwarfs 
can be found in e.g. \cite{bensby10b}.

(4) Perhaps the fault lies in systematic errors in [Fe/H].
There is a real sampling bias among the active groups
working in this area.  The initial results of 
the analyses of two microlensed MSTO bulge stars
by \citet{johnson07,johnson08} yielded 
surprisingly high metallicities.
Even Solar type dwarfs with such high [Fe/H] have 
complex spectra, with many blended and overlapping features,
and the photometry is not trustworthy in regions
with such high and variable reddening.
Cohen and Thompson, who lead the efforts at the Keck and Magellan
Observatories, believed that such a controversial issue could only
be finally settled on the basis of very high signal-to-noise spectra.
This meant
that they triggered target-of-opportunity observations
only for the brightest of the microlensed
bulge dwarfs, which tend to be those with the highest $A(max)$. 
The VLT group has observed events with a larger range
of brightness, and hence of $A(max)$.

In light of the sample bias between the VLT and the Magellan/Keck groups,
the relevant question is whether the [Fe/H] values determined
by the various groups involved are on a consistent scale.
Since the low magnification
events are in general fainter, one might
expect the resulting spectra to be 
on average of lower quality with lower signal-to-noise ratios.
Given that line crowding and blending make the
definition of the continuum in these spectra difficult,
this might produce a bias of underestimating
[Fe/H] in the required sense.

\cite{bensby09a} and \cite{bensby10b}  present comparisons
of multiple independent analyses by 
J.~Johnson, J.~G.~Cohen and T.~Bensby,
and collaborators of  HIRES or Mike spectra of six microlensed 
MSTO stars included in our sample.   The deduced [Fe/H] values are identical to within 
$\pm$0.10~dex in all cases.  
Very recently, the two lowest SNR spectrum from the
VLT sample, OGLE-2009-BLG-076S and MOA-2009-BLG-475, were also
analyzed by J.~Cohen, and even for these
very low SNR spectra the  derived [Fe/H] values were in
agreement to within the uncertainties, 
for the former being $-0.45\pm0.20$ vs. $-0.72\pm0.12$~dex and for the latter
$-0.49\pm0.20$ vs $-0.54\pm0.17$~dex.

We thus have established that there is
 good agreement in the derived [Fe/H] from detailed abundance
analyses when the three groups independently analyze the same spectrum of a
microlensed MSTO bulge star.  
The remaining issue is whether  [Fe/H] derived
from a spectrum of a microlensed
MSTO bulge dwarf is independent of the SNR within the range encompassed
by our sample of 16 microlensed bulge  MSTO stars. 
 T.~Bensby
has carried out a test on the high SNR Keck
spectrum of OGLE-2007-BLG349, degrading it to SNR $\sim 30$, and
finds that the deduced [Fe/H] changes by less than 0.05~dex.
We emphasize that errors in
[Fe/H] arising from different analyses or from the SNR
of the observations appear to be  too small
to be
the origin of the strong
correlation between $A(max)$ and [Fe/H] seen in Fig.~\ref{figure_max_mag}.

(5)  There are some biases in the detectability of a microlensing event that
depend on metallicity.  These arise because
the unlensed luminosity of a dwarf of
a given mass and age is a function of metallicity.
Tests with the Y$^2$ isochrones \citep{yi03} show that for a fixed
age (we adopt 10~Gyr), the mass of a star at the turnoff is
higher as metallicity increases and the turnoff becomes somewhat
fainter in $M_I$.   We adopt a Salpeter IMF, and compare two 
populations with this age
and with [Fe/H] between $-0.9$ and +0.6~dex.  While
for a population with a fixed total mass, the total number of
stars is significantly different in the two cases, the ratio of the 
the number of stars on the upper RGB selected in a fixed range
of $M_I$ to the
number of stars in the turnoff region is approximately constant.
Hence this cannot explain differences in the MDF between the bulge giants
and the bulge microlensed MSTO dwarfs nor lead to the correlation
between $A(max)$ and [Fe/H] seen in Fig.~\ref{figure_max_mag}.

(6) Perhaps the 
procedures by which microlensing events are identified
by the large surveys are biased
in some way.  This has been checked by  A.~Gould, who went through the entire
set of OGLE microlensing alerts from 2008, examining
each event, eliminating binaries, and redetermining $A(max)$ using
all available photometry when necessary.
He found that the number of microlensing alerts
as a function of $A(max)$, equivalent to $1/u_0$ for
$A(max) > 4$,  has the form expected for selection of a sample
unbiased in $A(max)$.  Fig.~\ref{figure_amp} shows the result, namely
there is a linear relationship between cumulative counts
and $u_0$  for $A(max) > 15$. This is the expected relation for 
uniform completeness in the OGLE microlensing survey 
over each of the three ranges of unlensed
source magnitude
considered. 
The  inset shows that the total number of events for $A(max) > 15$
in each bin
increases for fainter source stars, as expected, but this is
because there are more faint stars than bright ones.
Very high magnification events are very rare.
A table of updated $A(max)$ for the accepted 2008 OGLE events is
available as Table~2 (on-line version only).

While we believe that the source of the very strong
correlation seen in Figure~\ref{figure_max_mag} is some
bias in the sample of microlensed bulge MSTO dwarfs,
we have been unable to identify the source of the bias.
Every mechanism that we have thought of can be ruled out
with varying degrees of certainty.
Unfortunately,  until
the sample bias for the microlensed MSTO bulge dwarfs
is identified, 
the derived bulge-dwarf MDF and its comparison to the bulge-giant MDF
must be treated very cautiously.

Although suitable
high-magnification events are  rare and lining up the necessary
instruments/telescopes/clear weather at just the right time is difficult,
with the Keck, Magellan, and VLT observatories all quite interested
in this problem, the samples of MSTO bulge microlensed dwarfs
with high-dispersion spectra has risen rapidly, and will continue
to do so.  But, assuming that the correlation between
$A(max)$ of the event and [Fe/H] of the source star continues
to hold as the sample increases, what is really needed now
just as urgently as larger samples is a new insight
into what  is causing the very strong correlation
we have found between the maximum magnification of a
microlensing event for bulge MSTO stars and their metallicities
shown in Fig.~\ref{figure_max_mag}.

\acknowledgements

We are grateful to the many people  
who have worked to make the Keck Telescope and HIRES  
a reality and to operate and maintain the Keck Observatory. 
The authors wish to extend special thanks to those of Hawaiian ancestry
on whose sacred mountain we are privileged to be guests. 
Without their generous hospitality, none of the observations presented
herein would have been possible.  We thank Chris Hirata and Sterl Phinney
for helpful discussions.
J.G.C. thanks NSF grants AST-0507219 and AST-0908139  for partial
support. Work by A.G. was supported by NSF grant AST-0757888.
I.B.T. thanks NSF grant AST-0507325 for support.

{}

\clearpage

\begin{deluxetable}{lcrrr }
\tablenum{1}
\tablewidth{0pt}
\tablecaption{Data for Microlensed MSTO Bulge Stars \label{table_data}}
\tablehead{
\colhead{Name} & \colhead{Obs. Code\tablenotemark{a}} & \colhead{$A(max)$} &
\colhead{$A(obs)$\tablenotemark{b}} &
\colhead{[Fe/H]\tablenotemark{c}}  \\
\colhead{} & \colhead{} & \colhead{} &
\colhead{}    &   \colhead{(dex)} 
}
\startdata 
MACHO-1998-BLG-6 & K & 5 & 4 & $-0.22$ \\
MACHO-1999-BLG22 & K & 28 &  12 & $-0.35$ \\
MOA-2006-BLG99 & M & 515 & 110 & +0.36 \\
OGLE-2006-BLG265S & K & 210 & 135 & +0.56 \\
OGLE-2007-BLG349S & K & 450 & 400 & +0.56 \\
OGLE-2007-BLG514S & M & 1000 &  500 & +0.33 \\
OGLE-2008-BLG209 & M & 30 & 22 & $-0.33$ \\
MOA-2008-BLG-311 & M & 285 & 200 &  +0.26 \\
MOA-2008-BLG-310S & M & 380 & 313 & +0.42 \\
MOA-2009-BLG259 & K & 223 &  223 & +0.55 \\
OGLE-2009-BLG-076S & V & 68 & 48 & $-0.76$ \\
MOA-2009-BLG-493 & V & 150 & 123 & $-0.71$ \\
MOA-2009-BLG-133 & V & 74 & 35 & $-0.67$ \\
MOA-2009-BLG-475 & V & 62 & 48 & $-0.54$ \\
MOA-2009-BLG-489 & V & 103 & 103 & $-0.18$ \\
MOA-2008-BLG-456 & V & 77 & 47 & +0.12 \\   
\enddata
\tablenotetext{a}{M=Magellan, K=Keck, V=VLT.}
\tablenotetext{b}{Magnification at the time the spectroscopic observations were carried out.}
\tablenotetext{c}{References for each star are given in
\S\ref{section_sample} of the text.}
\end{deluxetable}

\clearpage

\begin{figure}
\epsscale{1.0}
\plotone{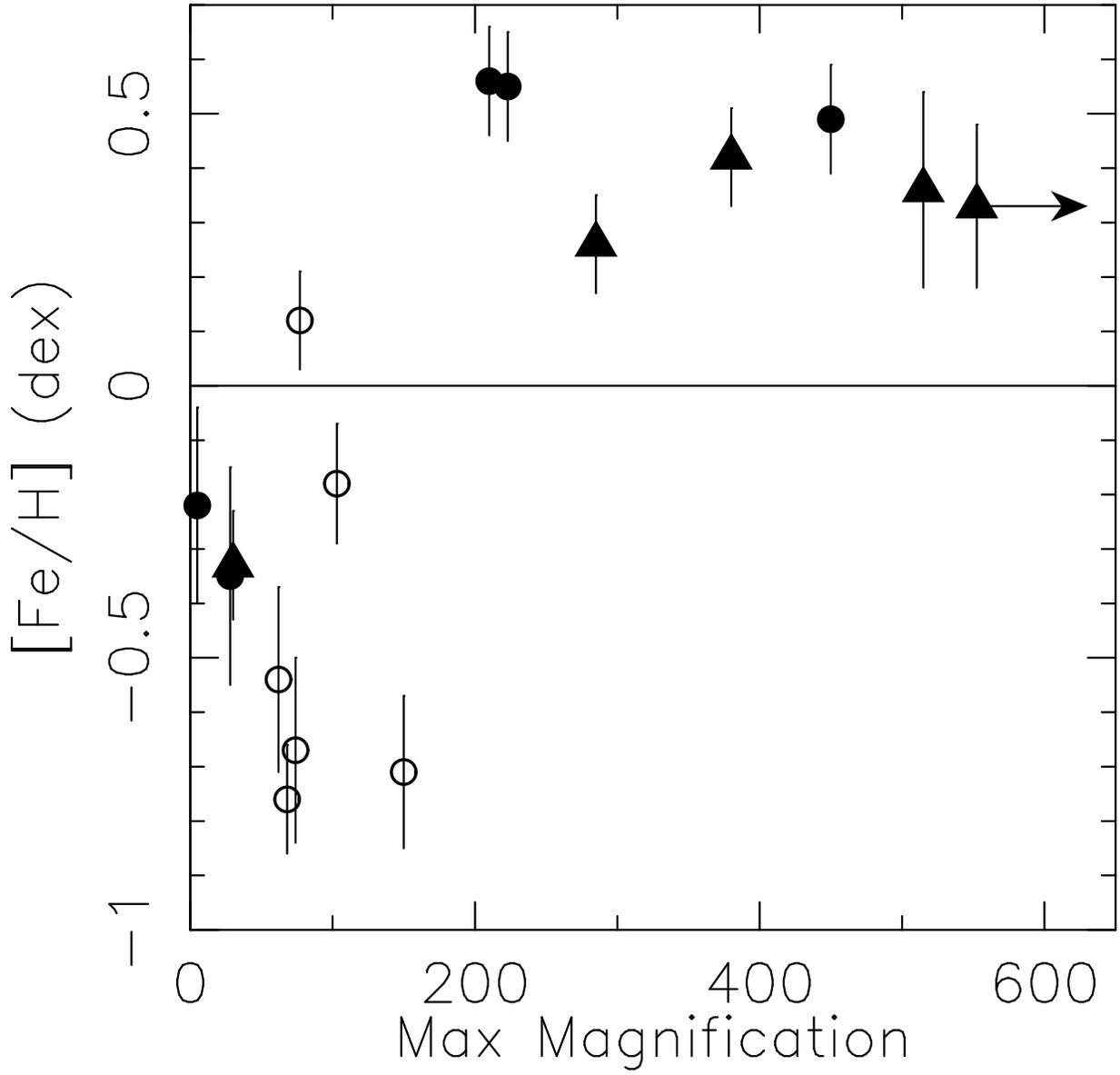}
\caption[]{[Fe/H] for the sample of 16
microlensed MSTO Galactic bulge stars
with detailed abundance analyses is shown as a function of the
maximum magnification achieved in each lensing event.  Filled circles denote
the Keck sample, filled triangles the Magellan sample,
and open circles denote those with VLT spectra.
\label{figure_max_mag}}
\end{figure}

\begin{figure}
\epsscale{1.0}
\plotone{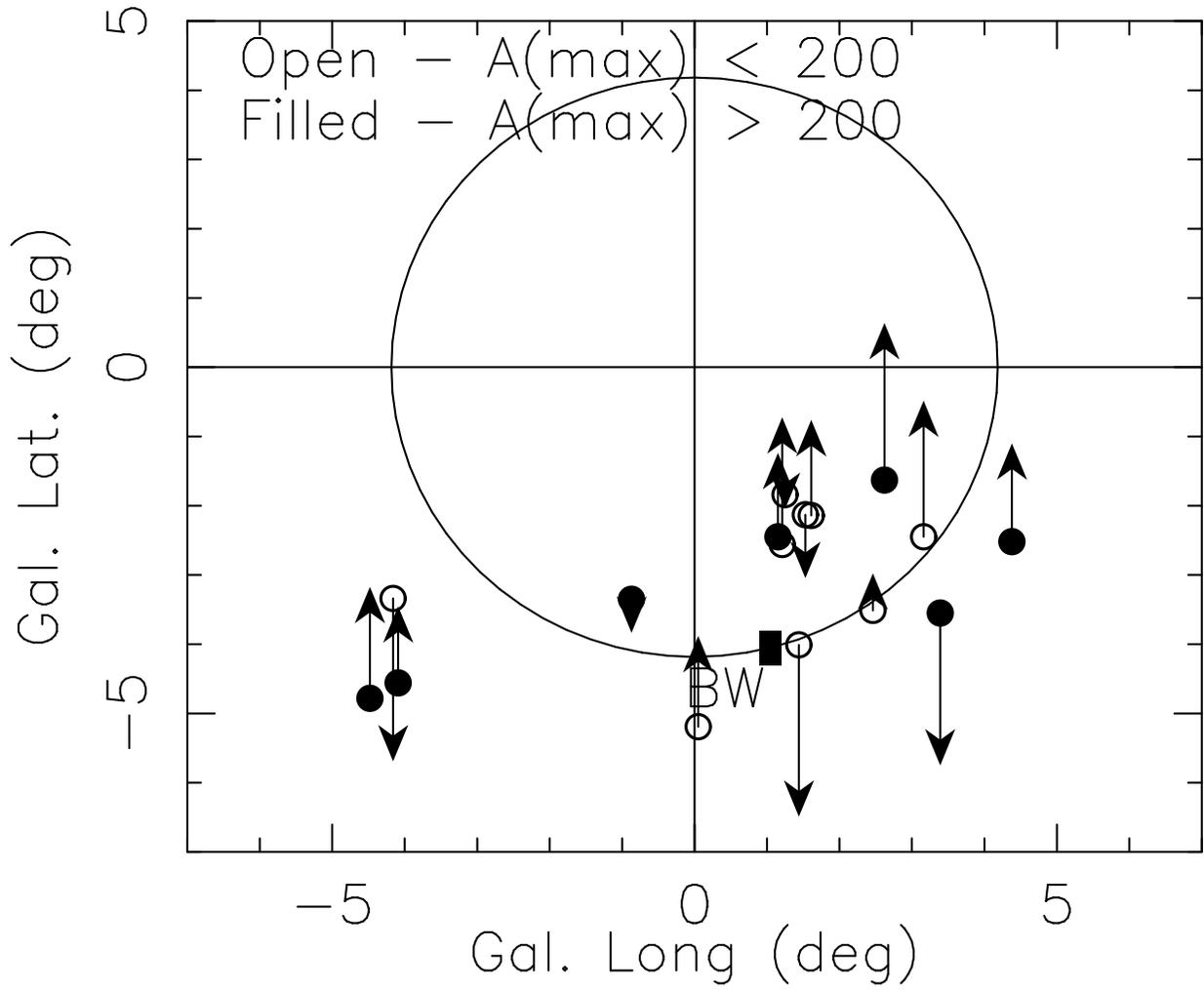}
\caption[]{The distribution in Galactic latitude and longitude of
the sample of 16 microlensed MSTO bulge stars. 
The  heliocentric radial velocity for
each star is indicated by an arrow, upward being positive, with a
scale of 70~km~s$^{-1}$ per degree.
Baade's Window is marked by the
filled rectangle, and its Galactocentric radius is indicated by a circle.
\label{figure_on_sky}}
\end{figure}

\begin{figure}
\epsscale{1.0}
\plotone{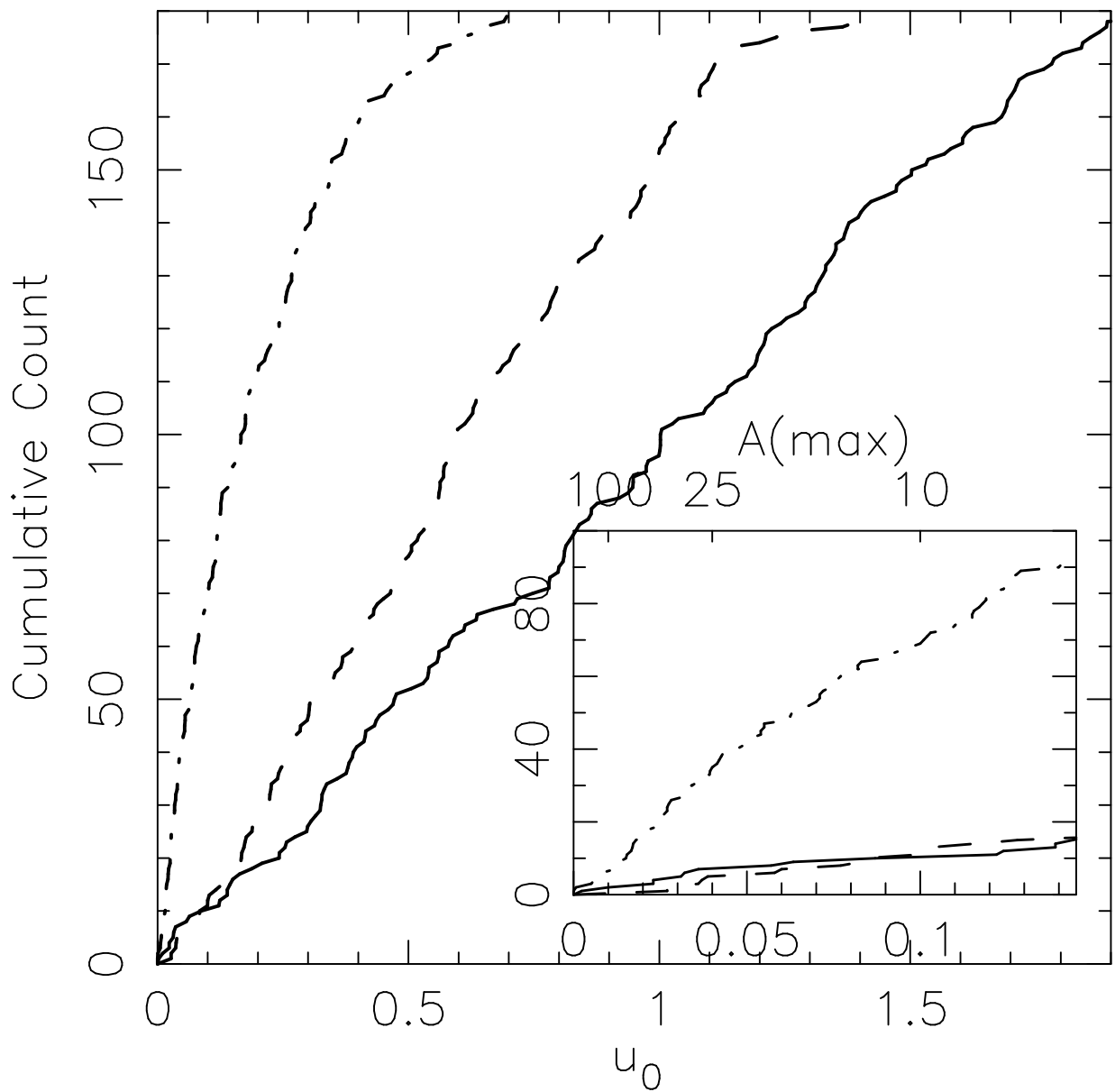}
\caption[]{Cumulative counts of microlensing events as a function of $u_0$ 
(equivalent to $1/A(max)$ for $A(max) > 4)$ for all OGLE alerts from 2008 that survived a check
by hand for validity.  Three different ranges of unlensed source
brightness are shown -
$I_S < 17.93$ (solid line), $17.93 < I_s < 19.28$ (dashed line), and 
$I_s > 19.28$~mag (dot-dashed line).
(green line); bin boundaries were chosen so that 1/3 of the sample
would be in each bin.
For microlensing event selection by the OGLE survey
to be independent of $A(max)$, this
relation should be a straight line, as is shown for high magnification events
in the inset. 
\label{figure_amp}}
\end{figure}

\clearpage

\begin{deluxetable}{lrrrr }
\tablenum{2}
\tablewidth{0pt}
\tablecaption{Check of 2008 OGLE Bulge Microlensing Alerts \label{table_alerts}}
\tablehead{
\colhead{OGLE} & \colhead{$u_0$} & \colhead{$t_E$} & \colhead{$I(source)$} &
\colhead{Code\tablenotemark{a}}  \\
\colhead{ID Number} & \colhead{} &
\colhead{(days)}    &   \colhead{(mag)} & \colhead{}
}
\startdata 
   2 &    0.4110  &     16.767  &     17.187  &    1  \\
   3 &    0.0123  &     99.967  &     21.882  &    2  \\
   4 &    0.1500  &     17.761  &     16.574  &    2  \\
   5 &    0.5054  &     17.630  &     14.734  &    2  \\
   6 &    0.2840  &     71.667  &     19.505  &    2  \\
   9 &    0.7880  &     28.962  &     18.509  &    1  \\
  10 &    0.0830  &     72.709  &     19.536  &    1  \\
  11 &    0.2990  &     44.934  &     17.364  &    1  \\
  13 &    0.0230  &     74.357  &     16.206  &    1  \\
  14 &    0.3030  &     23.270  &     18.468  &    1  \\
  15 &    1.1050  &      6.526  &     17.918  &    1  \\
  16 &    0.3320  &      6.370  &     17.833  &    1  \\
  18 &    0.0100  &     14.982  &     16.334  &    2  \\
  20 &    0.4320  &    103.531  &     17.046  &    1  \\
  21 &    1.2130  &     38.661  &     17.829  &    1  \\
  22 &    0.3270  &     15.585  &     17.503  &    1  \\
  25 &    0.2850  &     22.929  &     18.947  &    1  \\
  27 &    0.3650  &      5.295  &     18.528  &    1  \\
  29 &    0.1000  &      7.337  &     19.880  &    1  \\
  31 &    0.0870  &     25.390  &     16.524  &    1  \\
  33 &    0.4360  &     37.001  &     17.445  &    1  \\
  35 &    0.9480  &     29.282  &     15.884  &    1  \\
  36 &    0.3970  &     18.174  &     19.535  &    1  \\
  37 &    0.7100  &      8.447  &     18.439  &    1  \\
  38 &    0.1570  &     87.566  &     19.175  &    1  \\
  40 &    0.4070  &     16.810  &     18.913  &    1  \\
  41 &    0.0580  &     37.042  &     18.920  &    1  \\
  42 &    0.1220  &      4.540  &     16.582  &    1  \\
  43 &    1.3480  &      3.928  &     15.990  &    1  \\
  44 &    0.2100  &    101.967  &     19.193  &    1  \\
  45 &    0.8220  &     50.960  &     18.069  &    1  \\
  46 &    0.1750  &    115.225  &     19.337  &    1  \\
  47 &    0.2460  &     16.650  &     18.774  &    1  \\
  48 &    1.1330  &     14.773  &     17.813  &    1  \\
  49 &    1.6040  &     15.167  &     17.244  &    1  \\
  51 &    0.3670  &    286.473  &     19.498  &    1  \\
  54 &    0.0330  &     19.922  &     20.143  &    1  \\
  55 &    0.0550  &     38.869  &     20.026  &    1  \\
  56 &    0.2260  &      5.897  &     19.876  &    1  \\
  57 &    0.0510  &     10.161  &     19.448  &    1  \\
  58 &    0.4170  &     21.195  &     18.144  &    1  \\
  60 &    0.0630  &     20.655  &     19.544  &    1  \\
  61 &    0.1400  &      9.846  &     19.648  &    1  \\
  62 &    0.1260  &     47.132  &     20.635  &    1  \\
  63 &    0.4150  &      4.387  &     17.731  &    1  \\
  64 &    0.9720  &    144.140  &     18.528  &    1  \\
  65 &    1.4220  &     16.517  &     17.925  &    1  \\
  66 &    1.7170  &     25.112  &     15.386  &    1  \\
  67 &    0.0270  &     20.003  &     18.791  &    1  \\
  68 &    1.6940  &     52.685  &     15.391  &    1  \\
  69 &    0.0760  &     17.980  &     19.946  &    1  \\
  70 &    0.0060  &     27.053  &     21.391  &    2  \\
  71 &    0.7630  &     18.618  &     18.617  &    1  \\
  72 &    1.0010  &     10.290  &     17.897  &    1  \\
  73 &    1.1110  &      5.969  &     18.058  &    1  \\
  75 &    0.5890  &     49.688  &     18.467  &    1  \\
  76 &    0.4650  &     31.204  &     19.301  &    1  \\
  77 &    0.0270  &    147.016  &     22.163  &    1  \\
  79 &    1.0610  &     20.639  &     18.298  &    1  \\
  80 &    1.2420  &     13.690  &     17.926  &    1  \\
  81 &    0.6350  &     95.013  &     15.726  &    1  \\
  82 &    0.5790  &     14.229  &     16.769  &    1  \\
  83 &    0.0720  &     38.008  &     19.412  &    1  \\
  85 &    1.2110  &      7.514  &     17.468  &    1  \\
  86 &    0.1570  &    194.801  &     21.005  &    1  \\
  87 &    0.6560  &      3.265  &     19.139  &    1  \\
  88 &    0.3160  &     58.643  &     19.253  &    1  \\
  89 &    0.7800  &     25.708  &     17.749  &    1  \\
  90 &    0.4050  &     72.430  &     19.414  &    1  \\
  91 &    1.5680  &      7.871  &     17.494  &    1  \\
  92 &    0.7800  &     55.243  &     15.473  &    1  \\
  93 &    0.5610  &     15.265  &     18.020  &    1  \\
  94 &    0.2260  &     17.178  &     18.951  &    1  \\
  95 &    1.6820  &      8.903  &     17.121  &    1  \\
  96 &    0.5400  &    107.578  &     16.450  &    1  \\
  97 &    1.7120  &     33.313  &     16.937  &    1  \\
  99 &    0.1150  &     13.272  &     20.036  &    1  \\
 100 &    0.6820  &     22.663  &     19.106  &    1  \\
 101 &    1.2900  &     23.423  &     17.163  &    1  \\
 102 &    0.2840  &     61.710  &     17.978  &    1  \\
 103 &    0.8000  &      0.697  &     16.217  &    1  \\
 104 &    1.1220  &     12.682  &     18.542  &    1  \\
 105 &    0.1500  &     22.945  &     19.896  &    1  \\
 106 &    0.6960  &      9.439  &     19.493  &    1  \\
 107 &    0.3420  &      7.741  &     19.530  &    1  \\
 108 &    0.1600  &     56.427  &     19.629  &    1  \\
 109 &    0.4750  &     10.866  &     19.204  &    1  \\
 111 &    0.2470  &     58.643  &     20.274  &    1  \\
 112 &    0.3050  &     91.007  &     20.017  &    1  \\
 113 &    0.7480  &      9.762  &     17.875  &    1  \\
 114 &    1.3780  &     46.867  &     15.620  &    1  \\
 115 &    0.1500  &     32.691  &     20.783  &    1  \\
 116 &    1.0120  &     11.245  &     18.656  &    1  \\
 119 &    0.9780  &    167.630  &     17.819  &    1  \\
 120 &    0.4140  &     12.877  &     19.063  &    1  \\
 121 &    0.7200  &     49.192  &     18.884  &    1  \\
 123 &    0.0540  &     13.470  &     19.735  &    1  \\
 124 &    0.6420  &     20.399  &     18.811  &    1  \\
 126 &    0.6690  &     47.280  &     17.575  &    1  \\
 127 &    1.4100  &     26.572  &     17.809  &    1  \\
 128 &    1.0020  &      2.715  &     16.535  &    1  \\
 129 &    0.1880  &     84.009  &     18.446  &    1  \\
 130 &    0.5380  &      9.191  &     18.077  &    1  \\
 131 &    1.3700  &     15.818  &     16.989  &    1  \\
 132 &    0.1950  &     44.134  &     19.356  &    1  \\
 133 &    1.1000  &     37.567  &     18.203  &    1  \\
 134 &    1.3740  &      9.102  &     17.591  &    1  \\
 135 &    1.7010  &     61.954  &     14.646  &    1  \\
 136 &    0.2420  &     19.137  &     19.966  &    1  \\
 137 &    0.3510  &     71.293  &     19.034  &    1  \\
 138 &    0.1770  &     33.162  &     19.249  &    1  \\
 139 &    0.3020  &      9.248  &     19.028  &    1  \\
 140 &    0.0340  &     23.394  &     20.301  &    1  \\
 141 &    0.3220  &     16.292  &     19.559  &    1  \\
 142 &    0.3300  &      7.647  &     18.429  &    1  \\
 145 &    0.5590  &     29.933  &     19.466  &    1  \\
 147 &    1.2320  &     12.584  &     18.135  &    1  \\
 148 &    0.9840  &      7.901  &     18.073  &    1  \\
 149 &    0.3060  &      5.383  &     17.085  &    1  \\
 150 &    0.2390  &     24.258  &     19.102  &    1  \\
 151 &    0.3150  &      5.431  &     17.075  &    1  \\
 152 &    0.2730  &     26.361  &     20.249  &    1  \\
 153 &    0.8710  &      7.597  &     18.392  &    1  \\
 154 &    0.2720  &     52.374  &     18.985  &    1  \\
 155 &    0.0270  &     33.970  &     19.073  &    1  \\
 156 &    1.1740  &     28.587  &     17.191  &    1  \\
 158 &    0.4310  &      4.015  &     19.205  &    1  \\
 159 &    1.4490  &     38.588  &     17.171  &    1  \\
 160 &    0.7980  &      4.060  &     15.655  &    1  \\
 161 &    0.3690  &    137.506  &     18.680  &    1  \\
 162 &    1.1790  &      2.376  &     16.895  &    1  \\
 163 &    0.3760  &      4.904  &     17.284  &    1  \\
 164 &    0.8950  &     83.671  &     18.157  &    1  \\
 165 &    0.0770  &     39.656  &     20.601  &    1  \\
 166 &    0.0360  &     11.814  &     19.908  &    1  \\
 167 &    0.1160  &     17.843  &     19.092  &    1  \\
 168 &    0.5500  &      6.921  &     19.260  &    1  \\
 169 &    0.5460  &      7.926  &     19.496  &    1  \\
 170 &    0.3130  &    313.114  &     19.702  &    1  \\
 171 &    1.8910  &     20.235  &     13.839  &    1  \\
 172 &    0.5980  &      7.028  &     18.450  &    1  \\
 173 &    0.1660  &     21.493  &     19.432  &    1  \\
 174 &    1.5020  &     22.624  &     17.345  &    1  \\
 175 &    1.3960  &     20.018  &     17.363  &    1  \\
 176 &    0.4730  &      6.178  &     18.598  &    1  \\
 177 &    1.2240  &      2.778  &     17.785  &    1  \\
 178 &    0.9930  &      7.326  &     18.201  &    1  \\
 179 &    0.1410  &     69.365  &     21.101  &    1  \\
 180 &    0.2410  &      8.487  &     19.235  &    1  \\
 181 &    0.0490  &      6.601  &     19.966  &    1  \\
 182 &    0.0740  &     41.873  &     21.941  &    1  \\
 183 &    0.2080  &     14.114  &     16.389  &    1  \\
 185 &    0.7930  &     15.389  &     18.693  &    1  \\
 186 &    0.5700  &      3.512  &     19.107  &    1  \\
 188 &    0.0211  &      5.937  &     20.664  &    2  \\
 189 &    1.6120  &     16.042  &     16.641  &    1  \\
 190 &    0.9630  &     19.551  &     19.163  &    1  \\
 192 &    1.1510  &      8.149  &     17.622  &    1  \\
 193 &    0.2670  &     12.110  &     19.587  &    1  \\
 194 &    0.1920  &      4.073  &     19.282  &    1  \\
 195 &    0.4690  &     14.563  &     17.944  &    1  \\
 196 &    0.0570  &     22.087  &     17.289  &    1  \\
 197 &    0.1870  &     44.324  &     19.589  &    1  \\
 198 &    0.1670  &      4.632  &     19.067  &    1  \\
 199 &    0.0230  &     10.404  &     15.057  &    1  \\
 200 &    0.8550  &     24.919  &     17.975  &    1  \\
 201 &    0.3900  &      3.219  &     17.071  &    1  \\
 202 &    1.4720  &     65.209  &     17.920  &    1  \\
 203 &    1.0960  &     18.264  &     18.430  &    1  \\
 204 &    0.8650  &     43.871  &     16.877  &    1  \\
 205 &    1.6050  &     15.662  &     15.658  &    1  \\
 206 &    0.2650  &     40.297  &     18.865  &    1  \\
 207 &    0.7670  &     19.599  &     18.317  &    1  \\
 208 &    0.0310  &     24.011  &     16.705  &    1  \\
 209 &    0.0320  &     19.549  &     17.784  &    1  \\
 212 &    1.7820  &      7.416  &     17.146  &    1  \\
 213 &    0.5630  &     13.076  &     19.167  &    1  \\
 214 &    0.7830  &     12.514  &     18.715  &    1  \\
 215 &    0.1000  &      6.557  &     18.738  &    1  \\
 216 &    0.3910  &      9.651  &     18.244  &    1  \\
 217 &    1.3250  &     11.488  &     17.207  &    1  \\
 218 &    0.0500  &      5.712  &     19.596  &    2  \\
 220 &    0.5060  &      7.355  &     19.123  &    1  \\
 221 &    0.1680  &      9.516  &     19.284  &    1  \\
 222 &    0.9380  &     36.721  &     17.381  &    1  \\
 223 &    0.2980  &    211.089  &     18.897  &    1  \\
 224 &    1.3520  &     10.586  &     16.633  &    1  \\
 225 &    0.2600  &     38.254  &     18.798  &    1  \\
 226 &    0.5710  &      5.230  &     18.671  &    1  \\
 227 &    0.5820  &     15.501  &     17.839  &    1  \\
 228 &    1.6920  &     38.576  &     16.314  &    1  \\
 229 &    0.1390  &     53.994  &     17.688  &    1  \\
 230 &    0.4940  &      9.895  &     19.605  &    1  \\
 231 &    0.0362  &     42.592  &     20.249  &    2  \\
 232 &    0.3440  &      8.637  &     19.594  &    1  \\
 234 &    0.8220  &     13.720  &     18.828  &    1  \\
 235 &    0.8210  &      2.963  &     17.781  &    1  \\
 236 &    0.0820  &     10.354  &     20.400  &    1  \\
 237 &    0.3740  &     23.082  &     19.372  &    1  \\
 238 &    1.0320  &     18.433  &     18.822  &    1  \\
 239 &    0.4300  &      9.869  &     18.761  &    1  \\
 240 &    1.1060  &     21.793  &     18.567  &    1  \\
 241 &    0.3700  &      7.739  &     19.461  &    1  \\
 242 &    0.5890  &     13.436  &     17.170  &    1  \\
 244 &    1.7660  &     20.325  &     15.383  &    1  \\
 245 &    0.0248  &     26.516  &     19.684  &    2  \\
 246 &    0.3040  &      3.266  &     19.557  &    1  \\
 247 &    1.8440  &     12.598  &     16.436  &    1  \\
 248 &    0.1780  &     47.129  &     20.620  &    1  \\
 249 &    0.1155  &     15.442  &     19.990  &    2  \\
 250 &    1.2500  &     19.123  &     18.087  &    1  \\
 251 &    0.7010  &     46.523  &     19.183  &    1  \\
 252 &    0.4660  &     29.285  &     18.558  &    1  \\
 253 &    0.8740  &      2.942  &     19.000  &    1  \\
 254 &    0.0910  &      7.832  &     20.195  &    1  \\
 255 &    0.9030  &      1.228  &     18.056  &    1  \\
 257 &    0.0250  &      7.324  &     20.053  &    2  \\
 260 &    0.0830  &      8.336  &     18.372  &    1  \\
 261 &    0.0392  &     14.408  &     20.876  &    2  \\
 262 &    1.0490  &      9.563  &     18.518  &    1  \\
 264 &    0.0180  &      8.020  &     22.635  &    2  \\
 265 &    0.8140  &     38.742  &     17.428  &    1  \\
 266 &    1.0010  &     10.392  &     18.410  &    1  \\
 267 &    0.9430  &      9.989  &     19.174  &    1  \\
 268 &    0.6380  &     34.913  &     18.706  &    1  \\
 269 &    1.9930  &     14.415  &     15.287  &    1  \\
 270 &    0.0120  &    100.000  &     19.990  &    2  \\
 271 &    0.2800  &     28.039  &     20.149  &    1  \\
 272 &    0.0051  &     57.868  &     22.189  &    2  \\
 273 &    1.2080  &      6.524  &     17.731  &    1  \\
 274 &    1.2760  &      3.591  &     17.772  &    1  \\
 275 &    0.0620  &     57.587  &     21.741  &    1  \\
 277 &    0.1250  &      4.072  &     19.532  &    1  \\
 278 &    0.5570  &     18.850  &     20.108  &    1  \\
 279 &    0.0007  &     96.480  &     20.794  &    2  \\
 280 &    0.2420  &      7.106  &     20.133  &    1  \\
 281 &    0.1740  &      2.818  &     19.197  &    1  \\
 282 &    1.8770  &      7.477  &     16.471  &    1  \\
 283 &    0.2000  &     23.756  &     20.451  &    1  \\
 284 &    0.0740  &     37.999  &     20.745  &    1  \\
 285 &    0.2230  &     28.121  &     19.201  &    1  \\
 286 &    0.0340  &     36.494  &     19.611  &    1  \\
 287 &    0.4570  &     60.637  &     19.932  &    1  \\
 288 &    0.2560  &      8.024  &     19.364  &    1  \\
 289 &    0.9470  &      6.115  &     17.441  &    1  \\
 290 &    0.0023  &     16.226  &     16.908  &    2  \\
 291 &    0.8400  &      4.508  &     19.205  &    1  \\
 292 &    0.0322  &      3.335  &     20.367  &    2  \\
 293 &    0.1190  &    124.762  &     21.440  &    1  \\
 294 &    0.1660  &    140.097  &     19.906  &    1  \\
 295 &    0.9940  &     32.646  &     18.983  &    1  \\
 296 &    0.5780  &      5.492  &     19.024  &    1  \\
 297 &    0.5990  &     12.202  &     20.090  &    1  \\
 298 &    0.8890  &      7.782  &     18.940  &    1  \\
 299 &    0.0550  &     20.684  &     20.704  &    1  \\
 300 &    0.1100  &     22.513  &     20.416  &    1  \\
 301 &    0.1650  &     26.832  &     18.788  &    1  \\
 303 &    0.0220  &     35.902  &     19.684  &    2  \\
 304 &    0.4770  &     23.326  &     19.270  &    1  \\
 305 &    0.8410  &      2.272  &     17.537  &    1  \\
 306 &    0.3760  &     75.076  &     20.034  &    1  \\
 307 &    0.0387  &      4.074  &     19.264  &    2  \\
 308 &    0.0820  &     14.064  &     20.003  &    1  \\
 309 &    0.6360  &      7.583  &     19.172  &    1  \\
 310 &    0.3790  &     76.739  &     15.531  &    1  \\
 312 &    0.0540  &     32.006  &     19.816  &    1  \\
 313 &    0.3850  &     29.400  &     18.922  &    1  \\
 314 &    0.2680  &     45.988  &     20.380  &    1  \\
 315 &    1.4020  &      6.346  &     14.710  &    1  \\
 317 &    0.0700  &      7.056  &     19.506  &    1  \\
 318 &    0.2540  &     50.081  &     15.990  &    1  \\
 319 &    1.1290  &     34.808  &     18.337  &    1  \\
 321 &    0.5170  &      0.362  &     17.947  &    1  \\
 322 &    0.5000  &     27.832  &     18.802  &    1  \\
 323 &    0.7170  &      8.564  &     15.119  &    1  \\
 324 &    0.6120  &      7.817  &     18.454  &    1  \\
 325 &    0.3580  &     25.658  &     17.552  &    1  \\
 326 &    0.3440  &     32.978  &     19.689  &    1  \\
 327 &    0.0081  &     14.947  &     21.326  &    1  \\
 328 &    1.2010  &      2.712  &     17.690  &    1  \\
 329 &    0.6200  &      7.339  &     18.284  &    1  \\
 331 &    0.4950  &     56.046  &     19.733  &    1  \\
 332 &    1.3950  &     18.521  &     18.186  &    1  \\
 333 &    0.0360  &     10.454  &     15.053  &    2  \\
 334 &    0.4060  &      8.784  &     19.444  &    1  \\
 335 &    0.1180  &    104.358  &     19.564  &    1  \\
 336 &    0.1010  &     53.246  &     19.859  &    1  \\
 338 &    0.6100  &     11.048  &     16.518  &    1  \\
 339 &    0.3370  &     24.972  &     16.578  &    1  \\
 340 &    0.1240  &      6.607  &     16.653  &    1  \\
 341 &    0.0680  &     24.094  &     20.857  &    1  \\
 343 &    0.0710  &     26.663  &     20.284  &    1  \\
 345 &    1.1960  &     13.160  &     16.763  &    1  \\
 346 &    0.0433  &     42.426  &     20.730  &    2  \\
 349 &    0.0282  &     24.366  &     19.572  &    2  \\
 350 &    0.2910  &    117.256  &     19.724  &    1  \\
 351 &    1.3310  &      3.618  &     16.700  &    1  \\
 352 &    0.2670  &     36.215  &     19.784  &    1  \\
 354 &    0.1450  &     87.932  &     21.174  &    1  \\
 356 &    1.2540  &      9.402  &     17.811  &    1  \\
 357 &    0.9530  &      7.202  &     17.998  &    1  \\
 358 &    0.0151  &     63.449  &     22.228  &    2  \\
 359 &    0.0275  &      3.078  &     19.499  &    2  \\
 361 &    1.9900  &     30.017  &     16.378  &    1  \\
 362 &    1.4730  &      4.152  &     17.546  &    1  \\
 363 &    0.4190  &     21.120  &     19.737  &    1  \\
 364 &    0.2230  &     24.818  &     18.708  &    1  \\
 365 &    0.5820  &      1.253  &     19.238  &    1  \\
 367 &    0.0166  &      5.789  &     19.487  &    2  \\
 368 &    1.7330  &      4.157  &     14.766  &    1  \\
 369 &    0.2270  &     23.414  &     20.242  &    1  \\
 370 &    0.4510  &     32.244  &     20.039  &    1  \\
 371 &    0.1280  &     78.058  &     19.044  &    1  \\
 372 &    1.3190  &     12.852  &     13.819  &    1  \\
 373 &    0.4040  &      7.064  &     18.977  &    1  \\
 374 &    1.3670  &     40.375  &     15.980  &    1  \\
 376 &    1.7870  &     17.612  &     17.786  &    1  \\
 378 &    0.3250  &     14.065  &     15.811  &    1  \\
 379 &    0.0996  &     21.398  &     18.058  &    2  \\
 381 &    0.0394  &     56.161  &     21.651  &    2  \\
 382 &    0.1290  &     19.246  &     20.478  &    1  \\
 383 &    0.1030  &     26.611  &     18.358  &    1  \\
 384 &    0.0625  &     16.452  &     20.484  &    2  \\
 385 &    1.5030  &     11.916  &     16.891  &    1  \\
 386 &    1.0400  &     46.342  &     18.149  &    1  \\
 387 &    1.8590  &     10.853  &     17.293  &    1  \\
 388 &    0.1790  &      5.950  &     20.196  &    1  \\
 389 &    1.8930  &     60.328  &     16.388  &    1  \\
 390 &    0.3010  &     25.100  &     19.210  &    1  \\
 391 &    1.3410  &     26.816  &     17.490  &    1  \\
 392 &    0.0900  &     14.409  &     19.580  &    1  \\
 393 &    0.0360  &     25.254  &     18.996  &    1  \\
 394 &    0.3280  &     22.852  &     16.218  &    1  \\
 396 &    0.4380  &     21.745  &     18.573  &    1  \\
 397 &    0.1440  &      6.435  &     17.893  &    1  \\
 398 &    1.8040  &     19.059  &     17.302  &    1  \\
 399 &    1.7080  &      5.376  &     17.291  &    1  \\
 400 &    0.7430  &     15.860  &     18.922  &    1  \\
 401 &    0.9940  &      2.374  &     18.675  &    1  \\
 402 &    0.1030  &     22.956  &     20.509  &    1  \\
 403 &    0.8820  &      3.038  &     18.452  &    1  \\
 404 &    0.4720  &     47.022  &     17.707  &    1  \\
 405 &    1.0190  &      7.443  &     18.516  &    1  \\
 406 &    0.4430  &    135.727  &     17.821  &    1  \\
 407 &    0.2130  &     14.191  &     19.496  &    1  \\
 408 &    0.0970  &     70.366  &     21.145  &    1  \\
 409 &    0.1278  &      8.463  &     19.314  &    2  \\
 411 &    1.1170  &     10.462  &     18.160  &    1  \\
 412 &    0.2730  &     12.754  &     18.587  &    1  \\
 413 &    0.0420  &     71.227  &     19.946  &    1  \\
 414 &    0.3370  &     19.779  &     19.807  &    1  \\
 415 &    0.9630  &      7.086  &     18.541  &    1  \\
 416 &    0.1750  &     13.885  &     19.731  &    1  \\
 417 &    0.5380  &     30.552  &     17.669  &    1  \\
 418 &    1.0750  &     26.804  &     18.484  &    1  \\
 419 &    0.5170  &      1.691  &     19.546  &    1  \\
 420 &    1.2970  &      0.862  &     16.374  &    1  \\
 422 &    0.0940  &     51.855  &     19.534  &    1  \\
 424 &    0.7050  &      2.990  &     18.748  &    1  \\
 425 &    0.0651  &     13.534  &     20.034  &    2  \\
 427 &    0.1170  &     17.079  &     20.070  &    1  \\
 428 &    0.9990  &     20.815  &     17.593  &    1  \\
 429 &    1.5810  &     19.908  &     17.914  &    1  \\
 430 &    0.6040  &      5.960  &     19.479  &    1  \\
 431 &    0.1260  &      6.614  &     20.526  &    1  \\
 432 &    0.1830  &     12.875  &     20.486  &    1  \\
 433 &    0.3050  &    140.296  &     19.289  &    1  \\
 434 &    0.0470  &     16.235  &     19.382  &    1  \\
 435 &    0.2160  &      4.306  &     19.657  &    1  \\
 436 &    0.5980  &      6.403  &     19.272  &    1  \\
 437 &    1.3650  &     11.424  &     18.252  &    1  \\
 438 &    0.6880  &      9.784  &     19.104  &    1  \\
 439 &    0.3880  &     14.374  &     16.165  &    1  \\
 440 &    0.0270  &     25.729  &     22.025  &    1  \\
 441 &    0.5410  &     24.898  &     17.742  &    1  \\
 442 &    0.8100  &     35.378  &     15.994  &    1  \\
 443 &    0.3540  &      2.921  &     19.102  &    1  \\
 444 &    0.2610  &     11.301  &     19.845  &    1  \\
 445 &    1.6250  &      2.028  &     15.874  &    1  \\
 446 &    0.2400  &     39.482  &     19.356  &    1  \\
 447 &    0.2240  &     26.307  &     18.974  &    1  \\
 448 &    0.0130  &      7.294  &     20.117  &    2  \\
 450 &    0.9740  &      0.942  &     17.024  &    1  \\
 451 &    0.3410  &     28.792  &     19.658  &    1  \\
 452 &    0.2020  &     24.725  &     19.998  &    1  \\
 453 &    0.0840  &      9.360  &     18.521  &    1  \\
 454 &    1.3520  &      3.883  &     17.602  &    1  \\
 455 &    0.9590  &     38.684  &     18.833  &    1  \\
 457 &    0.6280  &     23.806  &     18.542  &    1  \\
 458 &    0.1890  &     13.473  &     18.837  &    1  \\
 459 &    0.2570  &     10.406  &     15.693  &    1  \\
 460 &    0.5600  &      8.696  &     16.219  &    1  \\
 461 &    0.7200  &      9.363  &     18.908  &    1  \\
 462 &    0.2270  &     27.770  &     18.824  &    1  \\
 463 &    0.2480  &    128.839  &     20.576  &    1  \\
 464 &    0.2640  &      7.622  &     19.199  &    1  \\
 465 &    1.0840  &     16.685  &     18.668  &    1  \\
 467 &    0.6900  &     39.481  &     19.354  &    1  \\
 468 &    0.5600  &     10.284  &     19.198  &    1  \\
 469 &    0.5910  &      9.910  &     19.029  &    1  \\
 470 &    0.2520  &     27.651  &     20.155  &    1  \\
 472 &    0.1730  &      9.497  &     20.063  &    1  \\
 473 &    1.6690  &      3.063  &     15.282  &    1  \\
 474 &    0.8360  &     33.763  &     19.083  &    1  \\
 475 &    0.1110  &     16.064  &     20.139  &    1  \\
 477 &    0.4500  &     11.825  &     18.931  &    1  \\
 478 &    0.2430  &     27.199  &     16.950  &    1  \\
 479 &    0.3818  &      0.364  &     17.172  &    2  \\
 480 &    0.1090  &     32.168  &     20.672  &    1  \\
 481 &    1.2000  &      6.392  &     18.085  &    1  \\
 482 &    0.5490  &     19.748  &     19.266  &    1  \\
 483 &    0.1200  &    108.671  &     19.948  &    1  \\
 484 &    0.6400  &     77.937  &     19.389  &    1  \\
 485 &    0.5290  &     36.986  &     16.370  &    1  \\
 486 &    0.2720  &     18.947  &     19.858  &    1  \\
 487 &    1.0100  &      9.668  &     18.666  &    1  \\
 488 &    0.3060  &     42.792  &     18.618  &    1  \\
 490 &    0.9790  &     15.707  &     18.674  &    1  \\
 491 &    0.6370  &     30.693  &     16.573  &    1  \\
 492 &    0.5590  &      6.773  &     19.181  &    1  \\
 494 &    0.9420  &     17.071  &     18.438  &    1  \\
 496 &    0.4000  &     16.050  &     20.085  &    1  \\
 497 &    0.7540  &      8.634  &     18.940  &    1  \\
 498 &    0.1720  &      4.420  &     19.054  &    1  \\
 499 &    0.5060  &     30.736  &     18.584  &    1  \\
 500 &    0.5290  &     28.069  &     19.398  &    1  \\
 502 &    0.7810  &     24.971  &     18.943  &    1  \\
 503 &    0.8080  &      5.603  &     14.719  &    1  \\
 504 &    1.1120  &      4.145  &     15.964  &    1  \\
 505 &    1.4830  &     17.909  &     16.925  &    1  \\
 507 &    0.4590  &      8.735  &     17.914  &    1  \\
 508 &    1.2910  &      3.361  &     17.270  &    1  \\
 509 &    0.0634  &      6.060  &     16.541  &    2  \\
 510 &    0.0600  &     21.642  &     19.211  &    1  \\
 512 &    0.1650  &     13.080  &     19.310  &    1  \\
 515 &    0.1240  &     28.342  &     20.778  &    1  \\
 517 &    1.0020  &     18.985  &     15.190  &    1  \\
 519 &    1.0210  &     19.208  &     18.040  &    1  \\
 520 &    0.0710  &    113.894  &     20.778  &    1  \\
 521 &    0.3680  &     10.199  &     19.135  &    1  \\
 523 &    0.8350  &      1.020  &     17.654  &    1  \\
 525 &    0.2970  &     13.054  &     17.243  &    1  \\
 526 &    0.1930  &     35.549  &     19.610  &    1  \\
 528 &    0.3950  &     14.820  &     18.551  &    1  \\
 529 &    0.5700  &      8.373  &     18.966  &    1  \\
 531 &    0.7760  &     35.459  &     18.594  &    1  \\
 532 &    0.8580  &      2.815  &     17.516  &    1  \\
 533 &    0.2580  &      6.458  &     19.620  &    1  \\
 534 &    0.8760  &     77.565  &     17.637  &    1  \\
 536 &    1.0380  &      1.942  &     16.819  &    1  \\
 537 &    0.7110  &      4.836  &     17.760  &    1  \\
 538 &    0.9990  &     35.077  &     18.821  &    1  \\
 539 &    0.4690  &     12.476  &     17.121  &    1  \\
 542 &    0.4880  &     32.895  &     19.134  &    1  \\
 543 &    0.0414  &      8.027  &     20.189  &    2  \\
 544 &    0.3240  &     51.769  &     17.781  &    1  \\
 546 &    0.2780  &      7.082  &     20.024  &    1  \\
 548 &    1.1940  &      1.326  &     17.926  &    1  \\
 550 &    0.6340  &      4.005  &     18.042  &    1  \\
 551 &    0.3140  &     87.720  &     19.684  &    1  \\
 552 &    0.2550  &      9.258  &     19.412  &    1  \\
 553 &    0.2720  &     16.275  &     19.630  &    1  \\
 554 &    0.7960  &      8.127  &     18.478  &    1  \\
 555 &    0.1633  &      4.431  &     17.312  &    2  \\
 556 &    0.0170  &      6.049  &     20.086  &    2  \\
 558 &    1.6880  &      5.801  &     15.644  &    1  \\
 560 &    0.5630  &     10.134  &     18.858  &    1  \\
 562 &    1.0800  &     20.369  &     18.143  &    1  \\
 563 &    1.0800  &     13.784  &     18.143  &    1  \\
 564 &    0.0154  &     48.033  &     20.759  &    2  \\
 565 &    0.5210  &     35.009  &     19.134  &    1  \\
 566 &    0.4180  &     13.212  &     20.080  &    1  \\
 568 &    0.1750  &     17.327  &     20.728  &    1  \\
 569 &    0.5310  &      6.221  &     19.187  &    1  \\
 571 &    1.5310  &      5.951  &     16.070  &    1  \\
 572 &    0.6500  &     21.002  &     19.597  &    1  \\
 573 &    0.1600  &     20.472  &     19.380  &    1  \\
 574 &    0.4760  &     21.279  &     16.261  &    1  \\
 575 &    1.0260  &      5.529  &     17.790  &    1  \\
 577 &    1.0880  &     19.076  &     17.110  &    1  \\
 579 &    1.3100  &      3.176  &     16.621  &    1  \\
 580 &    0.1610  &     12.775  &     18.580  &    1  \\
 582 &    0.6130  &     34.789  &     14.121  &    1  \\
 583 &    0.7820  &     19.729  &     17.293  &    1  \\
 585 &    1.8410  &      9.477  &     16.402  &    1  \\
 587 &    0.3450  &      8.151  &     19.522  &    1  \\
 589 &    0.4140  &      2.645  &     17.648  &    1  \\
 590 &    1.0020  &      4.988  &     17.476  &    1  \\
 591 &    0.5980  &     19.129  &     18.390  &    1  \\
 593 &    0.9480  &      3.298  &     15.812  &    1  \\
 595 &    0.9200  &     21.426  &     16.049  &    1  \\
 596 &    0.1150  &      9.811  &     19.672  &    2  \\
 597 &    0.3480  &     16.634  &     19.444  &    1  \\
 598 &    1.3320  &      5.988  &     16.158  &    1  \\
 599 &    0.8280  &      8.373  &     15.552  &    1  \\
 600 &    0.5570  &     30.395  &     16.637  &    1  \\
 601 &    0.1390  &     22.435  &     14.648  &    1  \\
 602 &    1.0050  &     18.815  &     17.858  &    1  \\
 604 &    0.3270  &     22.569  &     18.901  &    1  \\
 605 &    1.5360  &     14.926  &     16.404  &    1  \\
 607 &    0.3830  &     11.086  &     19.914  &    1  \\
 608 &    0.0073  &     17.533  &     21.042  &    2  \\
 609 &    0.8290  &     24.632  &     18.557  &    1  \\
 610 &    0.0251  &      0.172  &     21.514  &    2  \\
 611 &    1.0930  &      4.195  &     17.058  &    1  \\
 612 &    0.1750  &     13.667  &     20.034  &    1  \\
 614 &    0.4640  &     11.229  &     19.074  &    1  \\
 615 &    0.0207  &     94.549  &     22.290  &    2  \\
 616 &    0.1880  &     12.036  &     16.891  &    1  \\
 617 &    1.1900  &     18.426  &     14.581  &    1  \\
 618 &    0.1190  &     32.709  &     19.533  &    1  \\
 619 &    0.1530  &     10.256  &     18.821  &    1  \\
 620 &    0.0260  &     16.276  &     21.107  &    2  \\
 621 &    0.5610  &      6.337  &     16.530  &    1  \\
 622 &    0.0400  &     56.507  &     21.010  &    1  \\
 623 &    0.2950  &     10.212  &     19.571  &    1  \\
 624 &    0.8640  &     13.415  &     14.528  &    1  \\
 626 &    1.1380  &     24.520  &     16.455  &    1  \\
 627 &    0.2100  &      6.957  &     19.255  &    1  \\
 630 &    0.5320  &      8.272  &     19.031  &    1  \\
 631 &    0.0172  &     76.610  &     19.838  &    2  \\
 632 &    0.2200  &     49.643  &     19.713  &    1  \\
 634 &    0.0810  &     46.988  &     20.907  &    1  \\
 635 &    0.2420  &     15.499  &     17.635  &    1  \\
 636 &    0.0370  &     11.119  &     17.964  &    1  \\
 637 &    0.6670  &      4.239  &     18.118  &    1  \\
 638 &    0.0000  &     85.649  &     20.414  &    2  \\
 639 &    1.3130  &     10.793  &     17.478  &    1  \\
 640 &    0.9740  &      5.510  &     17.173  &    1  \\
 641 &    0.8110  &      6.714  &     14.634  &    1  \\
 644 &    0.2530  &     46.765  &     18.432  &    1  \\
 645 &    0.6290  &     25.406  &     18.706  &    1  \\
 646 &    0.0426  &     23.954  &     19.293  &    2  \\
 647 &    0.1020  &     91.987  &     20.506  &    1  \\
 651 &    0.2740  &      9.882  &     17.279  &    1  \\
 653 &    0.3970  &     36.205  &     17.838  &    1  \\
\enddata
\tablenotetext{a}{1 = original values unchanged, 2 = altered.  
Events for which $u_0$ could not be reliably determined or
estimated are not listed.}
\end{deluxetable}

\end{document}